\documentclass[aps,prl,twocolumn,superscriptaddress,nofootinbib,nobalancelastpage,showpacs,nobibnotes]{revtex4}
\usepackage{epsfig}

\begin{document}
\title
{Polarization precession in photon-photon encounters}

\author{R. F. Sawyer}\email{sawyer@vulcan.physics.ucsb.edu}
\affiliation{Department of Physics, University of California at
Santa Barbara, Santa Barbara, California 93106}

\begin{abstract}
We calculate the rate of precession of the direction of polarization of a photon traversing a sea of
plane-polarized photons moving in the opposed direction, where the interaction is the one-loop ``vacuum" 
Heisenberg-Euler coupling of four fields. Substantial precession can take place in a distance many
orders of magnitude shorter than the free path for photon-photon scattering, mediated by the same interaction.
We consider briefly the possibility of some interesting collective effects in the case in which instead of a 
particle and a sea, two seas are caused to collide.

\pacs{42.50.Xa}

\end{abstract}

\maketitle

\newcommand{\identity}{\:\mbox{\sf 1} \hspace{-0.37em} \mbox{\sf 1}\,}

\section{Introduction}

Some non-linear aspects of vacuum electrodynamics have been tested in 
experiments on Delbruck scattering \cite{ak1} , i. e. the scattering of a photon off
of the Coulomb field of a nucleus, and in photon splitting \cite{ak2}, also in the nuclear
Coulomb field. 

Essentially, these effects hinge on the one-loop effective Lagrangian density for processes in which four or more electromagnetic fields, of long wavelength compared to the electron Compton wavelength, come
together, as described by the Heisenberg-Euler interaction \cite{he} \cite{js}, the fourth order term of which is,

\begin{equation}
L_{I}= {2 \alpha ^2 \over 45 m^4}[(\bf E^2-B^2)^2+7(E \cdot B)^2 \rm]\,\,, 
\label{he}
\end{equation}
 where $\alpha$ is the fine structure constant and $m$ is the mass of the electron.
\footnote{We use units $\hbar=c=1$ throughout.}

The validity of the effective interaction term (\ref{he}), for long wave-length fields, can hardly be doubted.
Nonetheless, its confirmation in an actual photon-photon scattering experiment would be a milestone
of a kind.  Of course, if one puts in the numbers for photon-photon scattering itself,
the cross-section is far too small to be measured in an experiment on earth.
Indeed the ``light by light" scattering discussed in the very interesting experiment
reported in ref.\cite{slac} was the reaction $\gamma +\gamma \rightarrow e^+ +e^- $,
and does not test vacuum QED at the one loop level.
Nonetheless, the present note is inspired by the success of this group in doing a 
$\gamma +\gamma$ experiment.

The qualitative point is the following: Consider a cloud of photons $n_\gamma$
of average energy, $\omega_1$, with relatively small energy spread, and all moving
in more or less the same direction. We take the cloud to fill (briefly) some small volume and to have constant
number density within this volume. 
Now send in a beam of photons of energy $\omega$, where we shall think of 
$\omega>>\omega_1$ both for the purpose of keeping track of the beam photons,
and for the purposes of our later application. The mean interaction rate, $\Gamma_{\gamma,\gamma}$, of a beam photon,
due to photon collisions with the cloud photons is given by \cite{ib},
\begin{equation}
\Gamma_{\gamma,\gamma}=.014 \times \alpha^4 m^{-8} n_\gamma \, \omega^3 \omega_1^3\,\,.
\end{equation}
There exists another quantity, $\Gamma_{\rm pol}$, of dimensions ${\rm (time)}^{-1}$ which is of order,
\begin{equation}
\Gamma_{\rm pol}\sim \alpha^2 m^{-4} n_\gamma \,\omega \,\omega_1 \,\,.
\end{equation}
In all situations in which $\omega \, \omega_1<<m^2$ we see that $\Gamma_{\rm pol}$
is many orders of magnitude greater than $\Gamma_{\gamma,\gamma}$. The fundamental
remark of this paper is that if, in a Lorentz system in which the cloud and the beam are moving in opposed
directions, both beam and cloud have 100\% linear polarization, with an angle $\theta $ that is neither zero
nor $\pi/2$ between the two respective directions of polarization, then the polarization of a beam particle is
 substantially changed by the interaction (\ref{he}) during an interaction time $\Gamma_{\rm pol}^{-1}$.

Below, we address this problem first as a purely classical index-of-refraction effect; then we treat it from the
standpoint of the completely quantized system \footnote{It has been pointed out to us that Kotkin and Serbo \cite{ks} some time ago
derived index-of refraction expressions that are essentially the same as
given in our eqs.(\ref{disper}).
We have retained our own rather different derivations of the indices of refraction
both because they are different, and because they lay the 
groundwork for the speculative work on cloud-cloud
collisions in the latter part of this paper.}. Finally we shall discuss the possibility
that interesting non-perturbative effects might result from the complete solution of  the problem posed by the effective 
``forward" interaction term that gives the above effects when treated perturbatively.

\section{Classical treatment}
 We write the fields of a real vacuum plane wave, moving in the $\hat z$ direction
with energy $\omega_1$ and polarized in the direction $ \hat {\rm \bf p}$ as,
\begin{eqnarray}
{\rm \bf E_1}= \hat {\rm \bf p} F\sin [\omega_1 (z-t)+\phi]\,\, ,
\nonumber\\
{\rm \bf B_1}=\hat {\rm \bf z}\times {\rm \bf E_1}\,\,.~~~~~~~~~~~~
\label{classical}
\end{eqnarray}

Next we make the substitutions in (\ref{he}) of ${\rm \bf  B}\rightarrow {\rm \bf B }_1+{\rm \bf  B}$,
 ${\rm \bf  E}\rightarrow {\rm \bf E }_1+{\rm \bf  E}$, keeping the terms that are 
quadratic in ${\rm \bf  E}$ and ${\rm \bf  B}$, and averaging over the phase $\phi$, to obtain
\begin{eqnarray}
L_{eff}'= F^2 {2 \alpha ^2 \over 45 m^4}[4 \Bigr ({\rm \bf E}\cdot \hat {\rm \bf p}
+{\rm \bf B}\cdot (\hat {\rm \bf z} \times \hat {\rm \bf p}) \Bigr )^2
\nonumber\\
+7 ({\rm \bf B}\cdot \hat {\rm \bf p}+{\rm \bf E}\cdot (\hat {\rm \bf z} \times \hat {\rm \bf p}))^2] \,\,.~~~~~~~~~~~
\label{effective}
\end{eqnarray}

The strength constant $F$  is related to the number density, $n_\gamma$, of the cloud by
\begin{equation}
F=( 2 n_\gamma \omega_1)^{1/2}\,\,.
\end{equation}

With this connection, (\ref{effective}) is the effective Lagrangian for a beam photon, in a cloud of incoherent
photons of number density $n_\gamma$, all with the same plane-polarization $\hat {\rm \bf p}$, with roughly the 
same energies and traveling in roughly the same direction. The average over the phase $\phi$
is sufficient to capture all of the effects of incoherence that would have been included had
we begun with the superposition of fields of this incoherent cloud.

We take the total Lagrangian density, $L=({\rm \bf E}^2-{\rm \bf B}^2)/2+ L'_{eff}$ and 
derive the equations of motion. In Coulomb gauge, and resolving
the vector potential into components parallel and perpendicular to $\hat {\rm \bf p}$ we obtain simply,
\begin{eqnarray}
\partial _\mu \partial ^\mu {\rm \bf A}_{\parallel}={32\alpha ^2 \over 45 m^4}n_\gamma \omega^2 \omega_1 {\rm \bf A}_{\|}\,\,,
\nonumber\\
\,
\nonumber\\
\partial _\mu \partial ^\mu {\rm \bf A}_{\perp}={56 \alpha ^2 \over 45 m^4}n_\gamma \omega^2 \omega_1 {\rm \bf A}_{\perp}\,\,.
\label{fieldeqn}
\end{eqnarray}
To obtain (\ref{fieldeqn}) we substituted, on the RHS (after the equation of motion was derived by canonical
procedures),
\begin{equation}
 \partial_t \partial_ x  {\rm \bf  A}=\partial_t \partial_t  {\rm \bf  A}=\partial_x \partial_x {\rm \bf  A}=-\omega^2 {\rm \bf  A}\,\,,
\end{equation}
since we are calculating only a small change in the field of
the beam photon, where the unperturbed field is a plane wave moving in the $- \hat {\rm \bf z}$ direction. The dispersion relation to linear order in $\alpha^2$ for the beam is ,
\begin{eqnarray}
\omega_\parallel=k(1+4\beta) \,\,,
\nonumber\\
\omega_\perp=k(1+ 7 \beta) \,\,.
\label{disper}
\end{eqnarray} 
where $\beta=4 \alpha ^2 n_\gamma \omega_1/(45 m^4)$.
The polarization of the beam photon initially at an angle $\theta$ to $\hat {\rm \bf p}$ thus precesses
at a rate $\Delta \omega = 3\beta \omega $. Taking the initial polarization to be in the $\hat {\rm \bf x}$ direction,
the $\mid {\rm amplitude} \mid ^2$ of $\hat {\rm \bf x}$ polarization at later times is given by,
\begin{equation}
[P_x]_{\rm beam}=1-{1 \over 2}\sin^2 (2 \theta)[1- \cos (3 \beta \omega \, t)]\,\,.
\label{osc}
\end{equation}

\section{Quantum treatment}
Next we go back to the completely quantized theory, in order to confirm this result and to go
beyond. Consider the complete set of momentum states $\{q_i \}$ that are occupied
in the initial state, whether by cloud photons or by beam photons, retaining both polarization states for
each mode. We take $L_I$ of (\ref{he}) and 
truncate it by keeping the parts of the fields that contain only creation and annihilation operators for this
set of modes. The momentum-conserving processes described by this interaction
are just the forward scattering of beam photons from cloud photons since co-moving cloud
particles don't scatter from each other in this interaction.

To express the result, we introduce photon annihilation operators for the cloud modes, $b^x_j$, $b^y_j$, where $x$ and $y$ indicate the
polarization state and $j$ enumerates the set of momenta defined above, and corresponding operators
for the beam modes, $a^x_j$, $a^y_j$. 
We can then express the ``totally forward" interaction Lagrangian  in terms of the operators,
\begin{eqnarray}
\identity_j^{(b)} =(b^{(x)}_j)^\dagger b^{(x)}_j + (b^{(y)}_j)^\dagger b^{(y)}_j\,\,,
\nonumber\\
\zeta^{(1)}_j= (b^x_j)^\dagger b^y_j +(b^y_j)^\dagger b^x_j \,\,,
\nonumber\\
 \zeta^{(3)}_j=
(b^{(x)}_j)^\dagger b^{(x)}_j - (b^{(y)}_j)^\dagger b^{(y)}_j\,\,,
\nonumber\\
\identity_j^{(a)} =(a^{(x)}_j)^\dagger a^{(x)}_j + (a^{(y)}_j)^\dagger a^{(y)}_j\,\,,
\nonumber\\
  \tau^{(1)}_j=(a^x_j)^\dagger a^y_j + (a^y_j)^\dagger a^x_j \,\,,
\nonumber\\
\tau^{(3)}_j=
(a^{(x)}_j)^\dagger a^{(x)}_j - (a^{(y)}_j)^\dagger a^{(y)}_j \,\,.
\end{eqnarray} 

Note that the operators $\zeta^{(1),(3)}/2$, supplemented by an operator $\zeta^{(2)}/2$, which will not explicitly enter below,
obey angular momentum commutation rules, as do the operators $\tau^{(1),(3)}/2$. The effective ``forward" interaction Lagrangian
in terms of these operators is, 

\begin{eqnarray}
L_{\rm for}=-{3 \beta \omega  \over  {2\rm (Vol.)}}\sum_{j,m} [\zeta^{(1)}_j \tau^{(1)}_m + \zeta^{(3)}_j \tau^{(3)}_m
\nonumber\\
 -(11/3) \identity_j^{(a)}
 \identity_m^{(b)}] \,\,,
\label{forward}
\end{eqnarray}
where the indices $j$ and $m$ extend over the momentum states defined above.
As before, we take the cloud photons to be plane polarized in a direction that is in the $x-y$ plane
at an angle $\theta$ to the $x$ axis. Using $\langle \,\rangle$ to indicate the expectation value in this cloud
state, we have, 
\begin{eqnarray}
({\rm Vol.)}^{-1}\sum_j \langle \zeta_j^{(1)}\rangle =2 n_\gamma \,\sin (\theta) \cos (\theta)\,\,,
\nonumber\\
({\rm Vol.)}^{-1}\sum_j \langle \zeta_j^{(3)}\rangle=n_\gamma \,  [\cos^2 (\theta)-\sin ^2 (\theta)]\,\,,
\label{exp2}
\end{eqnarray}
and an effective Lagrangian for the evolution of the polarization of a single beam particle,
\begin{equation}
L_{\rm beam}=-{3\over 2} \beta \omega \Bigr [ 2 \sin (\theta) \cos (\theta) \sigma_1 +[\cos^2 (\theta)-\sin ^2 (\theta)] \sigma_3 \Bigr],
\label{Lbeam}
\end{equation}
where $\sigma _{1,3}$ are the Pauli matrices operating in the basis of states with (x,y) polarization.
The time advancement operator deriving from (\ref{Lbeam}) is just,

\begin{eqnarray}
U(t)=\exp[i L_{\rm beam} t] =\cos ({3\over 2} \beta \omega t) + i \Bigr [ 2 \sin (\theta) \cos (\theta) \sigma_1
\nonumber\\
 +[\cos^2 (\theta)-\sin ^2 (\theta)] \sigma_3 \Bigr] \sin ( {3\over 2} \beta \omega t)\,\,.~~~~~~~~~~~~~~
\end{eqnarray}
Now calculating,
\begin{equation}
[P_x]_{\rm beam} =\mid \langle \uparrow \mid U(t) \mid \uparrow \rangle \mid ^2\,\,,
\end{equation}
we regain the intensity oscillation formula (\ref{osc}) for the ${\hat {\rm \bf  x}}$  polarized direction.
Defining an oscillation length as $\lambda=(3 \beta \omega)^{-1}$ and expressing in terms of ordinary units
we have
\begin{equation}
\lambda=1.03 \times 10^{-8}\Bigr ( {E_{\rm crit} \over  E} \Bigr )^2 \Bigr ({ 1 {\rm MeV}\over \hbar  \omega})\,\, {\rm cm}
\end{equation} 
where $E_{\rm crit}=m^2c^3/e \hbar$ and $E$ is the rms electric field of the cloud.
In the $\omega_1=2.35 {\rm eV}$ laser used in the experiment
reported in ref.\cite{slac} the field strength was $ E / E_{\rm crit}\approx  1.5 \times 10^{-6} $. In this case taking $\hbar \omega = 1\,{\rm GeV}$ leads\footnote{Note that an energy of $1 {\rm GeV}$ in the lab corresponds to a center-of-mass energy that is below the threshold for pair production by more than a factor of ten.} to an oscillation length of $\approx 5 \,\,{\rm cm}$. Since the pulse length for the laser in ref. \cite{slac} is a fraction of a millimeter, an experiment appears not to be out of the question, although higher fields or longer pulse lengths, with a lower beam photon energy, might provide better possibilities. Just to compare the
above numbers with photon-photon cross-sections, we note
that for the above parameters, except now taking the laser field to fill all of space, the free path for scattering of the beam
photon would be approximately $10^9$ cm.
We should emphasize that in this calculation the laser served only to provide the high fields needed for
the polarization transformation. Incoherent cloud photons with the same rms field accomplish the same purpose.
The only coherence involved is the coherence of forward processes. 

\section{Collective effects}

We return to the fundamental assumption made in order to derive the above results, namely
that there is no dynamics involving the cloud operators ${\rm \bf E}_1$ and ${\rm \bf B_1}$,
beyond calculating their expectation value in the initial state as in 
(\ref{effective}) or (\ref{exp2}). While this is fairly clearly the correct procedure to obtain the index-of-refraction
effects with the time scale $\Gamma_{\rm pol}^{-1}$, we note the following:

1. Taking the initial polarizations to be mutually perpendicular, $\sin (2 \theta)=0$, there is no effect
predicted in the above treatment. There is a non-zero forward matrix element of $L_{\rm eff}$ that exchanges 
the polarization of a beam and a cloud photon, but it appears to contribute only at the $\Gamma_{\gamma \gamma}^{-1}$ 
time scale, since we have $\langle \zeta^{(1)} \rangle =0$ in the initial state.

2. If we consider a case in which the beam and cloud have comparable densities (which we shall
call the cloud-cloud case) then we should question the simplification, implicit in (\ref{classical}) or (\ref{exp2}), which
makes the operational Lagrangian quadratic in the dynamic variables. In this simplification we reduce
a product of four field operators, for example in the $B^4$ term in (\ref{he}), by, 
\begin{equation}
B^4\rightarrow B_2^2 \langle B_1^2 \rangle+\langle B_2^2 \rangle B_1^2\,\,
\label{replace}
\end{equation}
where $B_1$ and $B_2$ are the respective magnetic field operators for the two clouds. 

To clarify both of these points we briefly consider solutions to the Schrodinger equation for the 
system specified by the complete ``forward" interaction $L_{\rm for}$ in the cloud-cloud case.
In all that follows, we shall assume that the solutions of this equation might be relevant to physics,
despite the omission of almost all the modes of the field. Of course, these other modes will enter into the
wave function as time progresses, but it is our belief that their effects will average to zero through random
phases over the time that it takes for the forward effects add up to something. Provisionally accepting
this conclusion, our problem becomes very similar to that of some coupled discrete systems in which
evolution rates can be speeded up enormously over perturbative estimates, through collective effects \cite{rfs}.  

To make our point most simply we shift to circular polarization states. The forward interaction $L_{\rm for}$ gives a matrix
element for the angular-momentum-allowed transition in which a state of a positive helicity photon from one
bath and a positive helicity photon from the other bath makes a transition to a state with two photons of negative helicities. We can easily express $L_{\rm for}$
of (\ref{forward}) now in terms of operators $\xi^{(\pm)} ,\, \eta^{(\pm)}$ which act to raise and lower in the two dimensional helicity spaces of the two clouds, but defined such that $\xi^{(3)}$ and $\eta^{(3)}$ measure the spins in the $\pm \hat z$ direction for the photons in the respective baths, thus the negative of the helicity in the case of the down-moving photon.
We obtain,

\begin{eqnarray}
L_{\rm for}= {3\over 2} \beta \omega [ \xi ^{(+)} \eta ^{(-)}+\xi ^{(-)}\eta ^{(+)} 
\nonumber\\
 -(11/3) \identity ^{(a)} \identity ^{(b)} ] \, , 
\label{laser}
\end{eqnarray}

We have omitted the momentum subscripts and sums in (\ref{laser}), as though two laser beams
were colliding, so that only the four modes described by the two momenta and two polarizations enter. 
However, the essential results would be the same for incoherent light, subject to a caveat about the need for
correlation lengths to remain larger than the interaction region. 
 The dynamical problem that we now want to solve is one in which the initial
state $| I \rangle$ has $N_1$ up-moving photons and $N_2$ down-moving photons, all with helicity $+1$, where $N_{1,2}$ are
large. Thus we have $P_0=(\identity^{(a)}+\eta^{(3)})(\identity^{(b)}- \xi^{(3)})/4=
N_1 N_2$ in the initial state \footnote{We will use an
eigenstate of these number operators in the initial conditions, since what follows would look more 
complex if we took more physical
coherent states for the lasers.}, with zero values of the other three projectors. Then we 
calculate $| \Psi(t)\rangle=\exp [i L_{\rm for}t] \,| I \rangle$
and investigate, for example, how the expectation value  
of the $P_0$ projector develops with time. 
The evolution time is defined by the time required for this expectation to change by a 
macroscopic percentage from its original value of $N_1N_2$. We have solved the system
of $2 N_2 +1$ coupled differential equations for a case in which $N_1=3 N_2$ and for values of $N_2$ ranging from 
$2$ to $1024$. In this simulation we keep the densities fixed, scaling the volume to the number of particles.
The results indicate a characteristic time of $\Gamma_{\rm pol}^{-1}\log[N]$, where $N$ is of order $N_{1,2}$ \footnote{Of course, it is not possible to infer asymptotic behavior with certainty in a numerical experiment, especially when the behavior is
logarithmic.}. 
One's expectation based on the considerations discussed earlier had
been a characteristic time of order $\Gamma_{\gamma , \gamma}^{-1}$, since  
$\langle \eta^{\pm} \rangle=\langle \xi^{\pm} \rangle=0$ in the initial state. The speed-up that we find in the complete
calculation has been discussed previously in different contexts \cite{rfs}-\cite{fl}, and is attributable, more or less,
to the effects of entanglement in the many-particle system. \footnote{An analogue of our equations (\ref{fieldeqn}) is found in equations used in the evolution of neutrino clouds that play a role in the early universe before light element nucleosynthesis, and in the supernova atmosphere. These equations, again based on local four-field
couplings, had their beginnings with the synchronization results of ref.\cite{panta} and were developed further in ref.\cite{pastor}, and continued with much more literature that can be tracked from the review of ref.\cite{wong}. The issues raised in the present 
note with respect to the replacement (\ref{replace}) applies to these applications as well.} We can translate the result back into the plane polarization
picture to find that now, in the cloud-cloud problem, when the polarizations are initially orthogonal there is macroscopic
polarization exchange in the time $t \sim  \Gamma_{\rm pol}^{-1} \log [N]$.

Even with the
factor of $\log[N]$ this time is orders of magnitude less than $\Gamma_{\gamma , \gamma}^{-1}$ in all circumstances
of interest. Nevertheless, in the case of $\sin (2 \theta) \ne 0$, it appears to be correct to use the results of the reduction to a quadratic interaction (\ref{classical}), (\ref{exp2}), generically described in (\ref{replace}), for times
into the range $t \sim \Gamma_{\rm pol}^{-1}$, since the evidence
from the simulations discussed above is that the interesting collective effects evolve 
on a time-scale that is longer by a factor of order $\log[N]$ . Thus it appears that our estimates for the 
mixing rates addressed in the body of this paper are unaltered by the collective effects, although we have not performed numerical tests on the solutions for the full Lagrangian for cases $\sin (2 \theta) \ne 0$.

In summary, for the case of the beam and the cloud dynamics which was our primary focus, we have shown that a direct
experiment appears to be within range of present technology. This is an experiment that is very much worth doing, even
though one could be confident of the result. The case of two clouds, in our terminology, while very problematic
experimentally, to say the least, could test some fascinating properties of entangled many-body systems. 

\section{Acknowledgment} The author is indebted to Heidi Fearn for bringing ref. \cite{slac} to his attention.

\end{document}